\documentclass[aps,prl,floatfix,twocolumn,showpacs,reprint]{revtex4-1}
\usepackage{amsfonts}
\usepackage{mathrsfs}
\usepackage{amsmath}% needed for subequations
\usepackage{color}
\usepackage{graphicx}
\usepackage{bm}% bold maths
\usepackage{amssymb}
\usepackage{xspace}
\usepackage{epstopdf}
\usepackage{dcolumn}% Align table columns on decimal point
\usepackage{longtable}
\usepackage{subfigure}
\usepackage{multirow}
\usepackage[colorlinks=true, letterpaper=true, pdfstartview=FitV, linkcolor=blue, citecolor=blue, urlcolor=blue]{hyperref}

\newcommand{\sgn}{\text{sgn}}
\makeatletter

\newcommand{\Rmnum}[1]{\expandafter\@slowromancap\romannumeral #1@}
\makeatother
\begin{document}
\title{
       Time-Reversal-Invariant Topological Superconductivity and Majorana Kramers Pairs
       }
\author{Fan Zhang}\email{zhf@sas.upenn.edu}
\author{C. L. Kane}
\author{E. J. Mele}
\affiliation{Department of Physics and Astronomy, University of Pennsylvania, Philadelphia, PA 19104, USA}
\begin{abstract}
We propose a feasible route to engineer one and two dimensional time reversal invariant (TRI) topological superconductors (SC)
via proximity effects between nodeless $s_{\pm}$ wave iron-based SC and semiconductors with large Rashba spin-orbit interactions.
At the boundary of a TRI topological SC, there emerges a Kramers pair of Majorana edge (bound) states.
For a Josephson $\pi$-junction we predict a Majorana quartet that is protected by mirror symmetry and leads to a mirror fractional Josephson effect.
We analyze the evolution of the Majorana pair in Zeeman fields, as the SC undergoes a symmetry class change as well as topological phase transitions, providing an experimental signature in tunneling spectroscopy. We briefly discuss the realization of this mechanism in candidate materials
and the possibility of using $s$ and $d$ wave SCs and weak topological insulators.
\end{abstract}
\maketitle

{\color{cyan}{\indent{\em Introduction.}}}---
Broken symmetry and topological order are two fundamental themes of condensed matter physics.
The search for topological superconductors (SC)~\cite{pSC1,pSC2,MF} is fascinating, as gauge symmetries are spontaneously broken in the bulk and gapless Andreev bound states (ABS) can be topologically protected at order parameter defects, hosting Majorana fermions.
Majorana fermions are immune to local noise by virtue of their nonlocal topological nature and thus give hope for fault-tolerant quantum computing~\cite{TQC}.
The rise of topological superconductivity has been expedited by recent proposals~\cite{Kane-D,Fujimoto-D,Sau-D,Alicea-D,Lutchyn-D,Oreg-D} that hybridize ordinary SCs with helical materials, with the help of magnetic perturbations. Using proximity effects, electrons in a single helical band at the Fermi energy form conventional Cooper pairs, whose condensation realizes a spinless chiral $p$ wave SC in its weak pairing regime, {\em i.e.},
a topological SC with {\em broken} time reversal symmetry (class D). Unique signatures, including zero bias conductance peaks,
anomalous Fraunhofer patterns, and fractional Josephson effects, are starting to be observed in these systems~\cite{Kouwenhoven,Xu,Brinkman,DGG,Furdyna-FACJE}.
A completely distinct family (class DIII) of time reversal invariant (TRI) topological SCs was proposed based on a mathematical classification of Bogoliubov-de Gennes (BdG) Hamiltonians~\cite{Ryu,Kitaev,Qi,Teo,Qi1,Fu-3D,Sato,Metal-wire}.
${\rm Cu_xBi_2Se_3}$ \cite{Fu-3D} and Rashba bilayers~\cite{Nagaosa,Viola} are possible candidates,
however, it seems very challenging since exotic interactions are required and experimental observations remain controversial~\cite{Hor-3D,Wray-3D,Ando-3D,Kanigel,Jiang,Levy,LuLi,Peng}.

A more ambitious goal is to realize TRI topological SC {\em without} exotic electron-electron interactions in {\em absence} of Zeeman fields.
Here we propose a feasible route to utilize proximity effect devices which combine Rashba semiconductors (RS) and {\em nodeless} iron-based SCs.
Below its transition temperature the SC provides the RS a $s_{\pm}$ wave spin-singlet pairing potential that switches sign
between the $\Gamma$ and $M$ points~\cite{Hosono,Mazin,SC}. TRI topological SC is realized when the chemical potential is
adjusted to make the inner and outer Fermi surfaces feel pairing potentials with opposite signs.
At a boundary of the 2D (1D) TRI topological SC, a Kramers {\em pair} of Majorana edge (bound) states emerge as localized midgap states.
For a Josephson $\pi$-junction a Majorana {\em quartet} is protected by mirror symmetry~\cite{TMSC}, leading to a {\em mirror} fractional Josephson effect.
We analyze how the Majorana pair evolves in Zeeman fields, as the SC undergoes a symmetry class change
and topological phase transitions, providing an experimental signature in tunneling spectroscopy.
\begin{figure}[b]
{\scalebox{0.36} {\includegraphics*{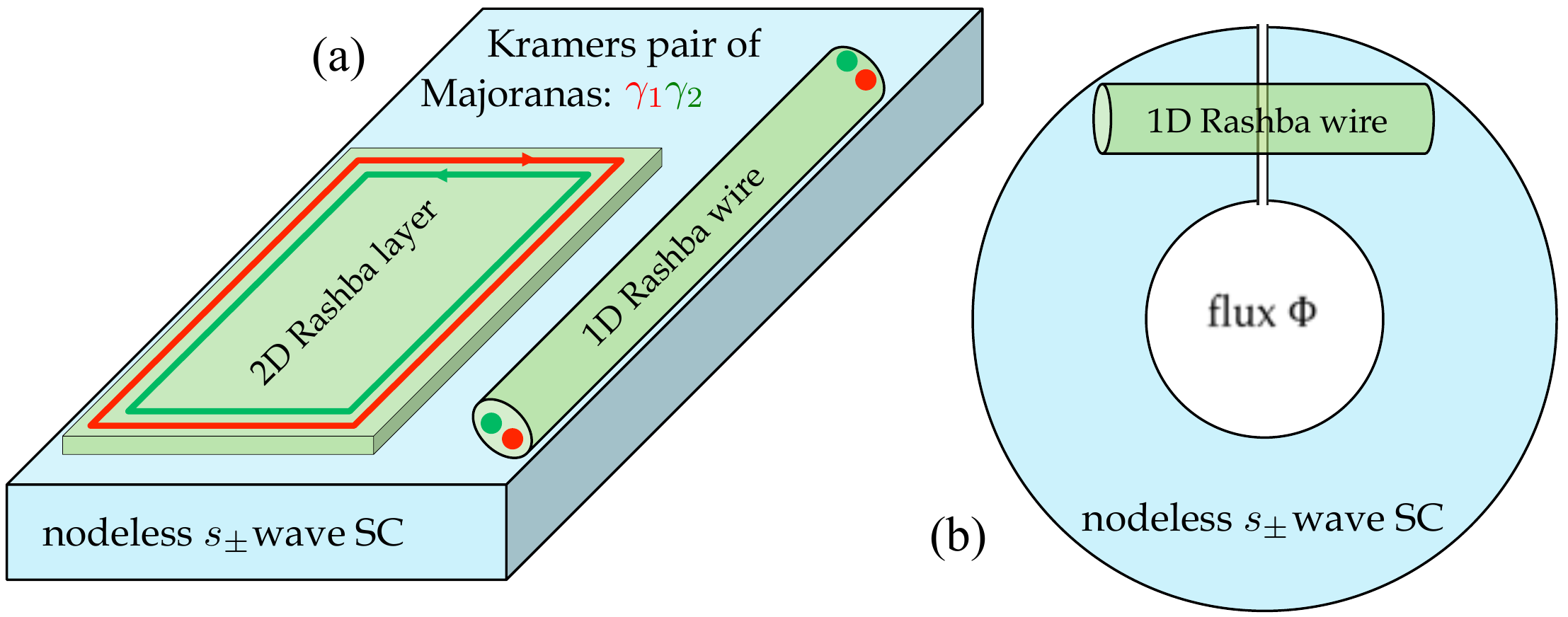}}}
\caption{\label{fig:fg1} {(a) Sketch of the proximity devices proposed in the text. A Majorana Kramers pair emerges at the boundary when the TRI SC becomes topological. (b) Sketch of the Josephson junction described in the text.}}
\end{figure}

{\color{cyan}{\indent{\em 2D TRI topological SC.}}}---
We first introduce a minimal model on a square lattice to characterize 2D TRI SC:
\begin{eqnarray}
\label{eqn:lattice}
H&=&-t\sum_{<ij>,\sigma}c^{\dag}_{i\sigma} c_{j\sigma}-i{\lambda_{\rm R}}\sum_{<ij>}c^{\dag}_{i\alpha}({\bm\sigma}_{\alpha\beta}\times \hat{{\bm d}}_{ij})_{z}c_{j\beta}\nonumber\\
&+&\Delta_0\sum_{i}(c^{\dag}_{i\uparrow}c^{\dag}_{i\downarrow}+\mbox{h.c.})+\Delta_1 \sum_{<ij>}(c^{\dag}_{i\uparrow}c^{\dag}_{j\downarrow} + \mbox{h.c.})\,.
\end{eqnarray}
Here $t$ is the nearest neighbor hopping and ${\bm \sigma}$ are the Pauli matrices of electron spin.
The second term arises from the Rashba spin-orbit interactions.
Note that $\hat{\bm d}_{ij}$ is a unit vector pointing from site $j$ to site $i$ and we assume $\lambda_{\rm R}>0$.
$\Delta_0$ and $\Delta_1$, induced by the proximity effect, lead to a combined $s_{\pm}$ wave pairing potential.
It is more convenient to write the BdG Hamiltonian:
\begin{eqnarray}
\label{eqn:BdG}
\mathcal{H}^{\rm BdG}_{\bm k}&=&[-2t(\cos k_x+\cos k_y)+h^{\rm R}_{\bm k}-\mu]\,\tau_z+\Delta_{\bm k}\,\tau_x\nonumber\\
{h}^{\rm R}_{\bm k}&=&2\lambda_{\rm R}(\sin k_x\,\sigma_y-\sin k_y\,\sigma_x) \nonumber\\
\Delta_{\bm k}&=&\Delta_0+2\Delta_1(\cos k_x+\cos k_y)\,,
\end{eqnarray}
where $\mu$ is the chemical potential and ${\bm \tau}$ are the Pauli matrices in Nambu particle-hole notation.
$\Delta_{\bm k}$ is a $s_{\pm}$ wave singlet pairing potential that switches signs between the zone center $\Gamma\;(0,0)$ and
the zone corner $M\;(\pi,\pi)$ when $0<|\Delta_0|<4\Delta_1$.
As we show in Fig.~\ref{fig:fg1}(a) and note below, $\Delta_{\bm k}$ could be provided
by a nodeless iron-based SC on which the Rashba layer is deposited.
$\mathcal{H}^{\rm BdG}_{\bm k}$ has time reversal (${\Theta}=-i\sigma_yK$) and
particle-hole (${\Xi}=\sigma_y\tau_yK$) symmetries. We obtain the energy dispersion
\begin{eqnarray}
E^{\rm BdG}_{\bm k}=\pm\sqrt{\left[2t(\cos k_x+\cos k_y)+\mu \pm \epsilon^{\rm R}_{\bm k}\right]^2+\Delta_{\bm k}^2}\,,
\end{eqnarray}
where $\epsilon^{\rm R}_{\bm k}=2\lambda_{\rm R}\sqrt{\sin^2 k_x+\sin^2 k_y}$ is the Rashba energy.
$\Delta_{\bm k}$ has a closed nodal line, {\em i.e.}, $\cos k_x+\cos k_y=-\Delta_0/(2\Delta_1)$, in the first Brillouin zone.
At the nodal line, $E^{\rm BdG}_{\bm k}=\pm\left(\mu-\epsilon_0 \pm \epsilon^{\rm R}_{\bm k}\right)$
with $\epsilon_0=t{\Delta_0}/{\Delta_1}$, and $\epsilon^{\rm R}_{\bm k}$ has the maxima
$\epsilon^{\rm R}_{max}=2\lambda_{\rm R}\sqrt{2-\Delta_0^2/(8\Delta_1^2)}$ and the minima
$\epsilon^{\rm R}_{min}=2\lambda_{\rm R}\sqrt{|\Delta_0/\Delta_1|-\Delta_0^2/(4\Delta_1^2)}$.

\begin{table}[b]
\caption{Summary of the $\mathbb{Z}_2$ classification of the hybrid TRI SC in class DIII.
$\epsilon_0$, $\epsilon^{\rm R}_{m}$, $\epsilon^{\rm R}_{min}$, and $\epsilon^{\rm R}_{max}$ are defined in the text.}
\newcommand\T{\rule{0pt}{3.1ex}}
\newcommand\B{\rule[-1.7ex]{0pt}{0pt}}
\centering
\begin{tabular}{c  c  c }
\hline\hline
 \;\, Phase \qquad & \qquad Two Dimension \qquad & \qquad One Dimension \T\\[3pt]
\hline
 \; $\nu=1$ & \qquad $|\mu-\epsilon_0|<\epsilon^{\rm R}_{min}$ & \qquad $|\mu-\epsilon_0|<\epsilon^{\rm R}_{m}$ \T \\[3pt]
  \; Nodal  & \qquad $\epsilon^{\rm R}_{min}\leq|\mu-\epsilon_0|\leq\epsilon^{\rm R}_{max}$ & \qquad
  $|\mu-\epsilon_0|=\epsilon^{\rm R}_{m}$ \T \\[3pt]
 \; $\nu=0$ & \qquad $|\mu-\epsilon_0|>\epsilon^{\rm R}_{max}$ & \qquad $|\mu-\epsilon_0|>\epsilon^{\rm R}_{m}$ \T \\[3pt]
\hline\hline
\end{tabular}
\label{table:one}
\end{table}

\begin{figure}[t]
{\scalebox{0.7} {\includegraphics*{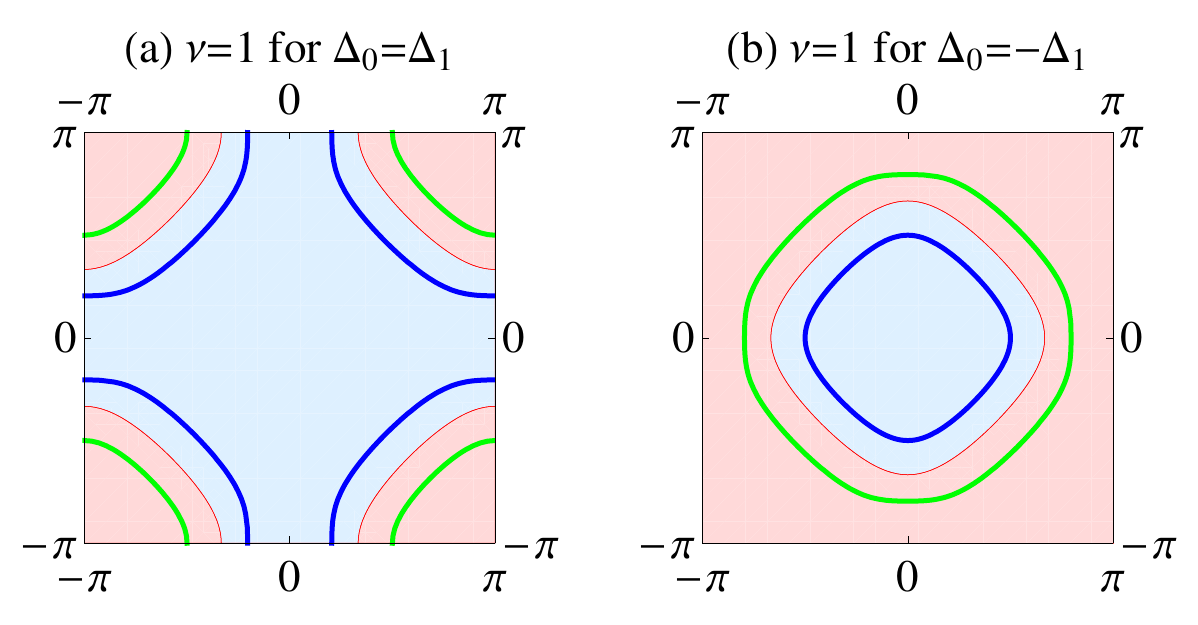}}}
{\scalebox{0.55} {\includegraphics*{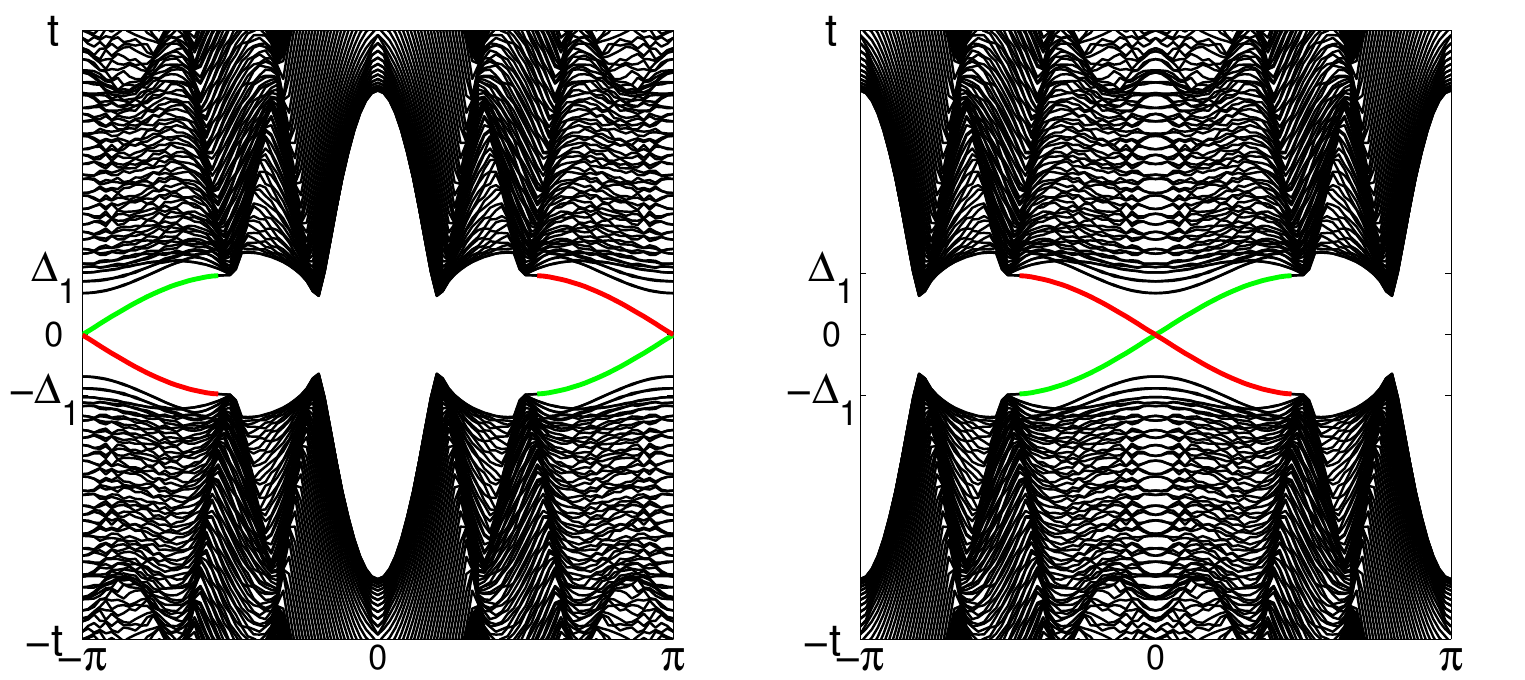}}}
\caption{\label{fig:fg2D} {Upper panels: the two Fermi surfaces (blue and green) of the single-particle bands for the $\nu=1$ state; the closed nodal line (red) of $\Delta_{\bm k}$, separating two regions in which $\Delta_{\bm k}$ has opposite signs.
Lower panel: BdG spectrum of a 2D ribbon as a function of $k$ for the $\nu=1$ state with $\mu=\epsilon_0$.
The red and green lines indicate the helical Majorana edge states.
We choose parameter values: $t=10$, $\lambda_{\rm R}=5$, and $|\Delta_0|=\Delta_1=2$. (a) $\Delta_0>0$ and (b) $\Delta_0<0$.}}
\end{figure}

In both 2D and 1D, the $\mathbb{Z}_2$ topological invariant \cite{Ryu,Kitaev,Qi,Teo} of a TRI SC is determined by
whether the pairing potential has a negative sign on odd number of Fermi surfaces each of which encloses a TRI momentum \cite{Qi}.
As shown in Fig.~\ref{fig:fg2D} and summarized in Table I \cite{supp}, the phase of the hybrid SC depends on the chemical potential $\mu$.
For the case of $\epsilon^{\rm R}_{min}\leq|\mu-\epsilon_0|\leq\epsilon^{\rm R}_{ max}$, $\mathcal{H}^{\rm BdG}_{\bm k}$ describes a nodal SC.
When $|\mu-\epsilon_0|>\epsilon^{\rm R}_{max}$, the SC is fully gapped but in the trivial ($\nu=0$) phase
since $\Delta_{\bm k}$ has the same sign on both Fermi circles.  When $|\mu-\epsilon_0|<\epsilon^{\rm R}_{min}$ is satisfied,
the pairing potential switches sign between the two Fermi circles,
and consequently the hybrid system realizes a TRI topological SC ($\nu=1$). The energy window for tuning the system into
the $\nu=1$ state has the size of $2\epsilon^{\rm R}_{min}$ with an optimized value $4\lambda_{\rm R}$ at $\Delta_0=\pm 2\Delta_1$.
For the $\nu=1$ state helical Majorana edge states emerge at the boundary, as shown in Fig.~\ref{fig:fg2D}.
This pair of Majorana edge states cross at $E=0$ and $k=\pi$($0$) for $\sgn(\Delta_0/\Delta_1)=+$($-$),
protected by time reversal and particle-hole symmetries.

{\color{cyan}{\indent{\em 1D TRI topological SC.}}}---
By turning off all the $k_y$ terms Eq.~(\ref{eqn:BdG}) models a 1D Rashba nanowire deposited on a nodeless $s_{\pm}$ wave SC.
When the two $s_{\pm}$ wave order parameters satisfy $|\Delta_0|<2\Delta_1$, the pairing potential switches sign
between the two TRI momenta $0$ and $\pi$.
In 1D, the closed nodal line of $\Delta_{\bm k}$ is shrunk to two nodes at $k=\pm\arccos({-\Delta_0}/{2\Delta_1})$.
At the nodes, the Rashba energy is $\epsilon^{\rm R}_{m}=2\lambda_{\rm R}\sqrt{1-\Delta_0^2/(4\Delta_1^2)}$,
and thus a proximity induced 1D TRI nodal SC is identified for $\mu=\epsilon_0\pm \epsilon^{\rm R}_{m}$.
When $|\mu-\epsilon_0|<\epsilon^{\rm R}_{m}$, a positive pairing is induced for the inner pair of Fermi points
while a negative pairing for the outer pair, realizing a 1D TRI topological SC.
In the case of $|\mu-\epsilon_0|>\epsilon^{\rm R}_{m}$, the hybrid system becomes a trivial SC
that is adiabatically connected to the vacuum state.

\begin{figure}[t]
{\scalebox{0.6} {\includegraphics*{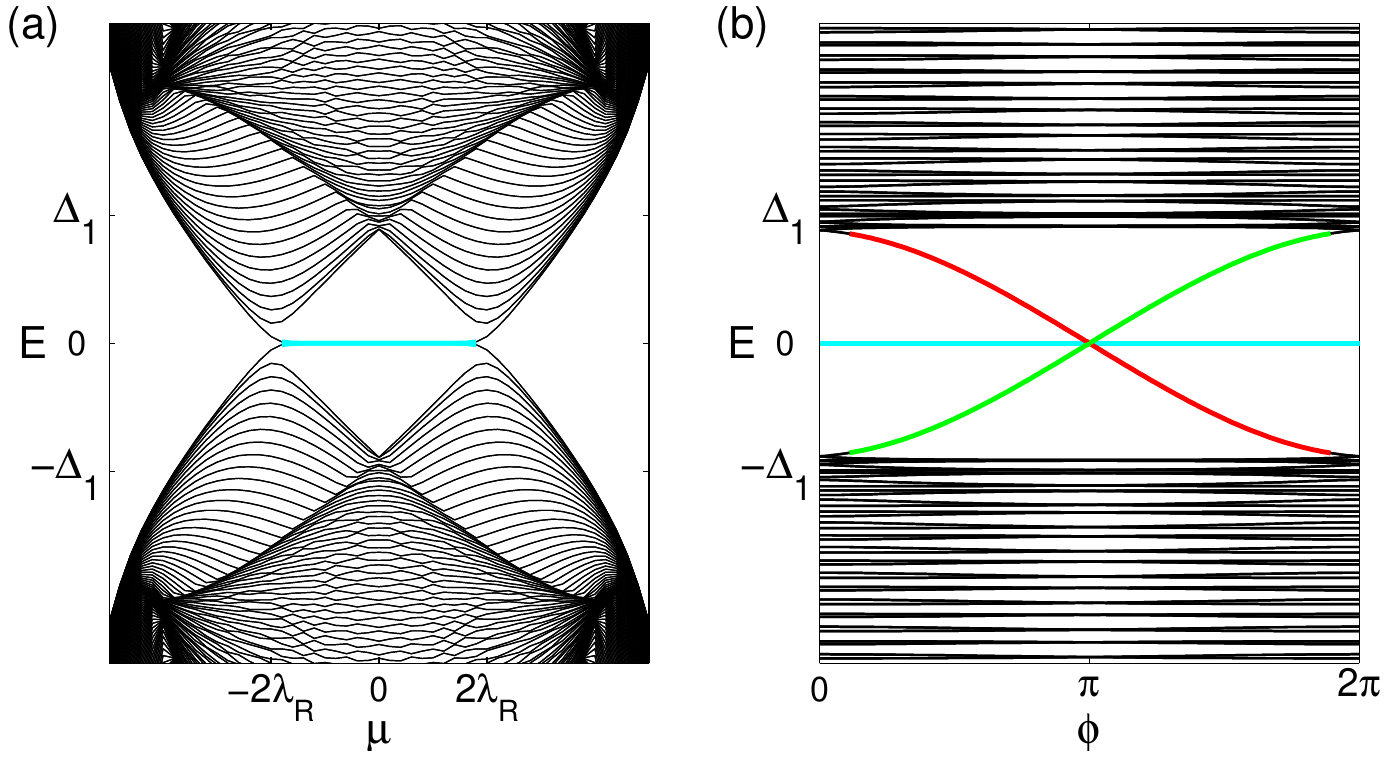}}}
\caption{\label{fig:fg3} {(a) BdG spectrum of the 1D TRI SC as a function of $\mu$. (b) Spectrum of ABSs in the junction as a function of $\phi$. The cyan lines are four-fold degenerate, denoting two Kramers pairs of MBSs at opposite ends. The red and green lines are doubly degenerate, denoting two pairs of ABSs in the junction. We choose parameter values: $t=10$, $\lambda_{\rm R}=5$, $\Delta_1=2$, and $\Delta_0=0$. $\mu=0$ is used in (b).}}
\end{figure}

At each end of a 1D TRI topological SC, there emerges a Kramers pair of Majorana bound states (MBS).
Without loss of generality, in the rest of this paper we will set $\Delta_0=0$ for the 1D case and
thus $|\mu|<2\lambda_{\rm R}$ is the criterion for the $\nu=1$ state, as shown in Fig.~\ref{fig:fg3}(a).
The cyan line denotes four degenerate MBSs independent of $\Delta_1$.
Further investigation of their wavefunctions shows that these four MBSs form two Kramers pairs localized at the opposite ends of nanowire.
This verifies our analytical results summarized in Table I.
We note that the two MBSs in each Kramers pair are time-reversal partners and form a special fermion level.
States with this level occupied and unoccupied are time-reversal partners,
and thus time-reversal symmetry acts like a supersymmetry that changes the fermion parity.

{\color{cyan}{\indent{\em Mirror Fractional Josephson effect.}}}---
Consider the linear Josephson junction in Fig.~\ref{fig:fg1}(b), in which a Rashba nanowire is deposited on a larger $s_{\pm}$ wave SC ring,
and the phase difference $\phi=(2e/\hbar)\Phi$ across the junction is controlled by the magnetic flux $\Phi$ through the ring.
Fig.~\ref{fig:fg3}(b) shows the spectrum of ABSs as a function of $\phi$ when the physical separation between the ends of SC ring is small.
The four-fold cyan line shows the appearance of a pair of MBSs at each end of the wire. The red and green lines, both doubly degenerate,
represent two pairs of ABSs in the junction. When $\phi\neq\pi$, time reversal symmetry is broken and a finite $\mu$ lifts the degeneracy.
However, protected by particle-hole, time reversal, and mirror ($\mathcal{M}_{y}=-i\sigma_{y}$) symmetries~\cite{TMSC},
their crossing at $E=0$ and $\phi=\pi$ is a four-fold degeneracy of special significance.
This $\pi$-junction is in sharp contrast with its $\mathbb{Z}_2$ counterpart in the class without mirror symmetry \cite{Teo,Lutchyn-D,Kane-M}.
Splitting the Majorana quartet at $\phi=\pi$ requires breaking mirror or time reversal symmetry~\cite{TMSC}.

To further understand this topological twist, we linearize our model (\ref{eqn:BdG}) near the Fermi energy:
\begin{eqnarray}\label{eqn:eff}
\mathcal{H}_{eff}&=&(vk_x\sigma_y-c)\,s\,\tau_z-\mu\,\tau_z+\mathcal{H}_{SC}\,,\\
\mathcal{H}_{SC}&=&\Delta\,s\,[\cos\frac{\phi}{2}\,\tau_x+\sgn(x)\sin\frac{\phi}{2}\,\tau_y]\,,
\end{eqnarray}
where $s=\pm$ denote the inner and outer bands with opposite spin helicities and $c$ ($0<c<\mu$) lifts their degeneracy.
We find the ABS dispersions $\epsilon_{s}(\phi)=\Delta\cos({\phi}/{2})$.
Note that the perfect normal state transmission and the independence on $\mu$, $c$ and $s$ are artifacts of the simplified model.
We can define Bogoliubov operators $\Gamma_{s\pm}$ that satisfy $\Gamma_s\equiv\Gamma_{s+}=\Gamma^{\dag}_{s-}$
because of the particle-hole symmetry. The low-energy Hamiltonian is thus ${H}=\sum_{s}\epsilon_{s}(\phi)(\Gamma_{s}^{\dag}\Gamma_{s}-\frac{1}{2})
=2i\sum_{s}\epsilon_{s}(\phi)\gamma_{s}\eta_{s}$
where $\gamma_{s}=(\Gamma_{s}^{\dag}+\Gamma_{s})/2$ and
$\eta_{s}=i(\Gamma_{s}^{\dag}-\Gamma_{s})/2$ are the Majorana operators.
The mirror symmetry allows one to label the bands with the eigenvalues of $\sigma_y$.
For each $s$, $\Gamma_{s}^{\dag}\Gamma_{s}=0,1$ distinguishes two states with different mirror eigenvalues
and coupling them requires a process that changes $\Gamma_{s}^{\dag}\Gamma_{s}$.
Due to the Cooper pairing, the total charge is not conserved.
However, the fermion parity $\Gamma_{s}^{\dag}\Gamma_{s}$ is conserved, as the mirror symmetry does not allow scattering between the two bands.
This {\em mirror fermion parity} conservation forbids to mix the four ABSs in the junction
and therefore protects their crossing at zero energy.

The phase difference $\phi$ acts like a defect and parameterizes the mirror fermion parity pump.
Although Eq.~(\ref{eqn:eff}) is invariant under $\delta\Phi=h/2e$, the global Hamiltonian is physically distinct.
When a flux $h/2e$ is threaded through the SC ring, $\phi$ is advanced by $2\pi$,
$\gamma_{s}\rightarrow\gamma_{s}$ while $\eta_{s}\rightarrow-\eta_{s}$,
and a unit of fermion parity is transferred between the two bands resolving the fermion parity anomaly for an individual band.
In response to the phase change, the populated ABSs carry supercurrents $I_{s\pm}=\pm I_{s}$,
whereas the states in the continuum has negligible contributions.
We obtain $I_{s}=({e}/{\hbar}){\partial\epsilon_{s}}/{\partial\phi}\sim\sin({\phi}/{2})$,
which is maximized at $\phi=\pi$ in sharp contrast to the $\nu=0$ (conventional) case.
In the absence of mirror symmetry breaking, there is no transition among $I_{s\pm}$,
signaling a mirror fractional Josephson effect with $4\pi$ periodicity.

\begin{figure}[t]
{\scalebox{0.51} {\includegraphics*{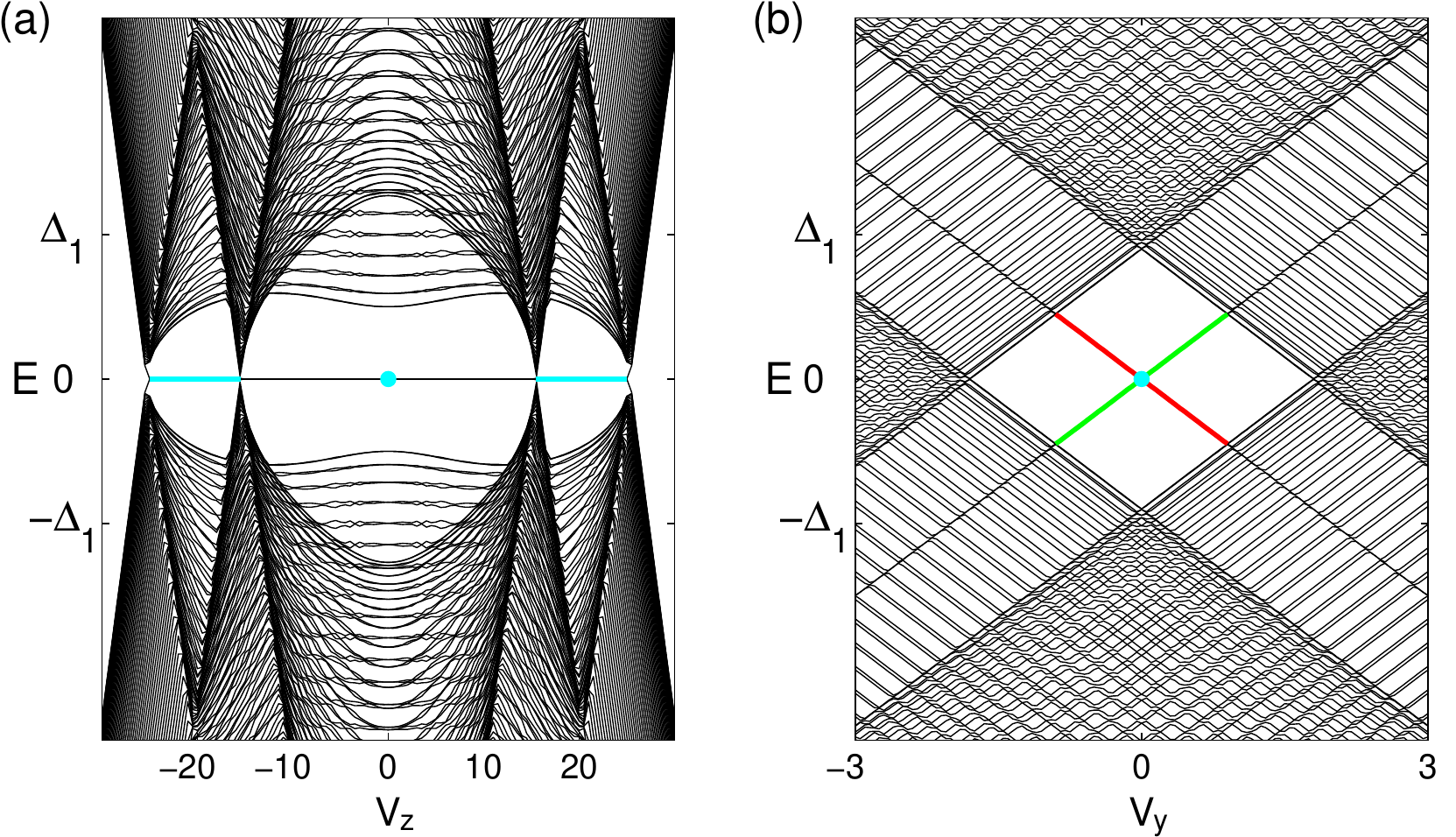}}}
\caption{\label{fig:fg4} {Evolution of a MBS pair in Zeeman fields. (a) In a $\sigma_z$ ($\sigma_x$) field; (b) in a $\sigma_y$ field.
The cyan dots (lines) have two-fold (no) degeneracy, indicating the appearance of a MBS pair (single MBS) at the end of SC.
The black lines at zero energy are doubly degenerate.
The red and green lines indicate the Zeeman splitting of two zero energy states with opposite fermion parity.
We choose parameter values: $\mu=-5$ in (a), $\mu=0$ in (b), and others are the same as in Fig.~\ref{fig:fg3}.}}
\end{figure}

{\color{cyan}{\indent{\em Evolution of Majorana pair in Zeeman field.}}}---
When one helical band is removed from the Fermi energy, a Rashba nanowire proximity coupled to a $s$ wave SC is a topological SC
with broken time reversal symmetry, supporting a single MBS at each end.
Since only one band is present at the Fermi energy this also occurs even if the SC is $s_{\pm}$ wave.
Realizing such a topological phase requires $\mu$ and a Zeeman field $V_z\sigma_z$ (or $V_z\sigma_x$) to satisfy $4\Delta_1^2+(|\mu|-2t)^2<V_z^2<4\Delta_1^2+(|\mu|+2t)^2$,
provided that the hybrid SC remains fully gapped ($\mu^2\neq4\lambda_{\rm R}^2+V_z^2$).
The latter condition is guaranteed for the TRI $\nu=1$ phase as it satisfies $|\mu|<2\lambda_{\rm R}$.
It is thus intriguing to investigate how the Majorana pair evolves in the Zeeman field,
as the bulk SC undergoes a symmetry class change and topological phase transitions.
Fig.~\ref{fig:fg4}(a) shows the evolution in the case of $\mu\neq0$. When $V_z$ is turned on, without gap closing,
the topological SC in the TRI class becomes a trivial SC in the class without time reversal symmetry.
Two topological phase transitions occur at $V_z^2=4\Delta_1^2+(|\mu|\pm2t)^2$ where the gap closes.
The $\nu=1$ state in the new class is realized between the transitions. At one end, as $V_z$ is tuned up,
one MBS disappears at the first transition while the other persists in the $\nu=1$ state and enters the bulk continuum at the second transition.
For the special case $\mu=0$, the two transitions merge into one and the new $\nu=1$ state does not appear.

As implied by the zero energy black lines in Fig.~\ref{fig:fg4}(a),
it seems baffling that the Majorana pair is robust against the Zeeman field before the first transition occurs.
We emphasize that these Majorana modes are not topologically protected because the bulk SC is trivial in the new class.
However, their robustness can be understood using the effective model described by Eq.~(\ref{eqn:eff}) with
\begin{eqnarray}\label{Eq:tbc}
\mathcal{H}_{SC}=\Delta\,s\,\tau_x\,\theta(a-|x|)+\bar{\Delta}\,\tau_x\,\theta(|x|-a)\,,
\end{eqnarray}
where $x=\pm a$ are the locations of the two ends of SC and $\bar{\Delta}\rightarrow\infty$ reflects the infinity mass of vacuum.
Eq.~(\ref{Eq:tbc}) incorporates the correct boundary condition \cite{TISS} of TRI topological SC. Solving the boundary problem,
we find that a pair of MBSs at one end forms two states with $\sigma_y=\tau_y=\pm1$
while the other pair at the opposite end forms two states with $\sigma_y=-\tau_y=\pm1$.
Clearly, a $\sigma_{x}$ or $\sigma_{z}$ Zeeman field only couples states with opposite $\sigma_y$ flavors at opposite ends,
whose wavefunction overlap decays exponentially on the SC length.
Consequently, none of the Majorana states can be passivated away from zero energy until the first topological phase transition.~\cite{BdI}
However, as shown in Fig.~\ref{fig:fg4}(b), a small $\sigma_y$ field can Zeeman split the two zero energy states
with opposite fermion parity at each end, revealing the topological triviality of bulk SC. The $\sigma_y$ field also closes the SC gap.

The evolution of a Majorana pair in a Zeeman field provides a smoking gun for the identification of a TRI topological SC
in tunneling spectroscopy. In a single-channel quantum point contact, the Majorana pair induces resonant Andreev reflection producing
a quantized zero bias conductance peak \cite{RAR,PC,review} of $4e^2/h$.
This peak persists for small $\sigma_{x}$ and $\sigma_{z}$ fields, reduces to $2e^2/h$ after the first topological quantum phase transition,
and disappears at the second one. In contrast, a small $\sigma_{y}$ field not only Zeeman splits but also reduces the peak.

{\color{cyan}{\indent{\em Discussions.}}}---
Unlike a $d_{xy}$ wave SC, where it is impossible to induce a full gap at the Fermi surface centered at a TRI momentum,
a $s_{\pm}$ or $d_{x^2-y^2}$ wave SC allows the pairing potentials on Fermi surfaces
centered at $(0,0)$ and $(\pi,\pi)$ or at $(0,\pi)$ and $(\pi,0)$ to have opposite signs.
Thus it is possible to build a 2D TRI topological SC by hybridizing a weak topological insulator
with two surface states and a $s_{\pm}$ or $d_{x^2-y^2}$ wave SC~\cite{supp}.
Although rotational symmetry of RS makes a hybrid 2D $d_{x^2-y^2}$ wave SC nodal,
it is possible to use them to engineer a 1D TRI topological SC,
as suggested by Wong and Law \cite{Law}.
One might wonder whether it is possible to utilize a pure $s$ wave SC.
We note that this seems implausible~\cite{supp}.

Our proposal is experimentally feasible, since the necessary ingredients and required technologies are all well established. It has been widely accepted that, at least for iron pnictides, there are a large family of nodeless $s_{\pm}$ wave SCs~\cite{Hosono,Mazin,SC},
though their pairing mechanism is still under lively debate.
An iron-pnictide with closer electron and hole Fermi surfaces is preferred.
$\Delta_{0,1}$ can be adjusted by doping or changing materials.
Iron-pnictides consist of square layers that are stacked in tetragonal structures and
it is better to choose RS with cubic or tetragonal structures, {\em e.g.}, Au, Ag, and Pb~\cite{direction}.
Small lattice incommensurability may blur and effectively broaden the induced pair potential in RS,
which may be helpful as long as the hybrid SC remains nodeless.

Our proposal for a realization of TRI topological SCs in 1D and 2D constitutes six critical advances:
(i) there is no need for magnetic perturbations or exotic interactions, simplifying the experimental setup;
(ii) the SC gap can reach more than $15$ meV, raising the critical temperature of topological SC;
(iii) the time reversal symmetry likely provides an Anderson's theorem to mitigate the role of bulk disorder,
making the superconductivity more robust;
(iv) irrelevant bands are absent and higher subbands play no role~\cite{irrelevant}, allowing large tunability in feasible materials;
(v) the chemical potential is not necessarily close to the band degeneracy point of RS,
where the electron density is low and the disorder effect is large;
and (vi) the Majorana states are stable to pair fluctuations, as long as the fluctuations neither close the gap nor change the pairing sign at any Fermi surface.
Our work also provides a way to use the presence of Majorana fermions to test the pairing symmetry of unconventional SCs.

{\color{cyan}\indent{\em Acknowledgements.}}---This work has been supported by DARPA grant SPAWAR N66001-11-1-4110.
CLK has been supported by NSF grant DMR 0906175 and a Simons Investigator award from the Simons Foundation.


\begin{thebibliography}{100}
\bibitem{BdI}
Alternatively, one can think that the Hamiltonian with a $\sigma_x$ or $\sigma_z$ Zeeman field has an effective time-reversal symmetry ($\Theta=K$)
and thus belongs to the BDI symmetry class with an integer classification.
$\tau_y\sigma_y=\Theta\Xi$ is the new chiral symmetry operator and our state is topological ($\nu=2$) in this new symmetry class.
This result is consistent with the analysis in the main text.
\bibitem{direction}
 For the 2D case, the thin film is best grown along $(001)$ direction to utilize the square symmetry.
 $(111)$ thin films with hexagonal symmetry may require a large $\lambda_R$. For the 1D case, either growing direction works.
\bibitem{irrelevant}
 The bulk bands~\cite{Kane-D} or the band removed from Fermi energy~\cite{Sau-D,Alicea-D,Lutchyn-D,Oreg-D} plays negative roles in realizing a class D topological SC experimentally. There are no such problems in our case. Furthermore, the topological criterion in class DIII only cares about the sign of pairing potential at Fermi surfaces, implying that higher subbands not present at Fermi energy play no role.
\end{thebibliography}
\end{document}

\title{
       Supplementary Materials
       }
\author{Fan Zhang}\email{zhf@sas.upenn.edu}
\author{E. J. Mele}
\author{C. L. Kane}
\affiliation{Department of Physics and Astronomy, University of Pennsylvania, Philadelphia, PA 19104, USA}
%\begin{abstract}
%\end{abstract}
%\date{\today}
\pacs{74.45.+c, 71.70.Ej, 71.10.Pm, 74.78.Na, 73.22.Gk, 74.78.Fk}
% 74.45.+c Proximity effects; Andreev reflection; SN and SNS junctions
% 71.10.Pm Fermions in reduced dimensions (anyons, composite fermions, Luttinger liquid, etc.)
% 74.78.Fk Multilayers, superlattices, heterostructures
% 71.70.Ej spin-orbital
% 74.78.Na Mesoscopic and nanoscale systems
% 73.22.Gk broken symmetry state
\maketitle

The main messages of the present work is that one can proximity-couple a Rashba layer (wire) and a nodeless $s_{\pm}$ wave superconductor
to build a two (one) dimensional time reversal invariant topological superconductor with the emergence of a Majorana Kramers pair on the boundary.
Such a topological superconductor is fundamentally different with the commonly discussed class D topological superconductor [ref.~1-15] with a single Majorana mode on the boundary.

In the discussion part of the main text, we mentioned that (i) in principle it is also possible to utilize a weak topological insulator to achieve the goal
and that (ii) it is likely impossible to utilize a pure $s$ wave superconductor.
Although these two points are not related to the main messages of the present work,
we would like to explain them further in a simple way for specialists who are also interested in them.

\subsection{I. Using a weak topological insulator}

\begin{figure}[h]
{\scalebox{0.5} {\includegraphics*{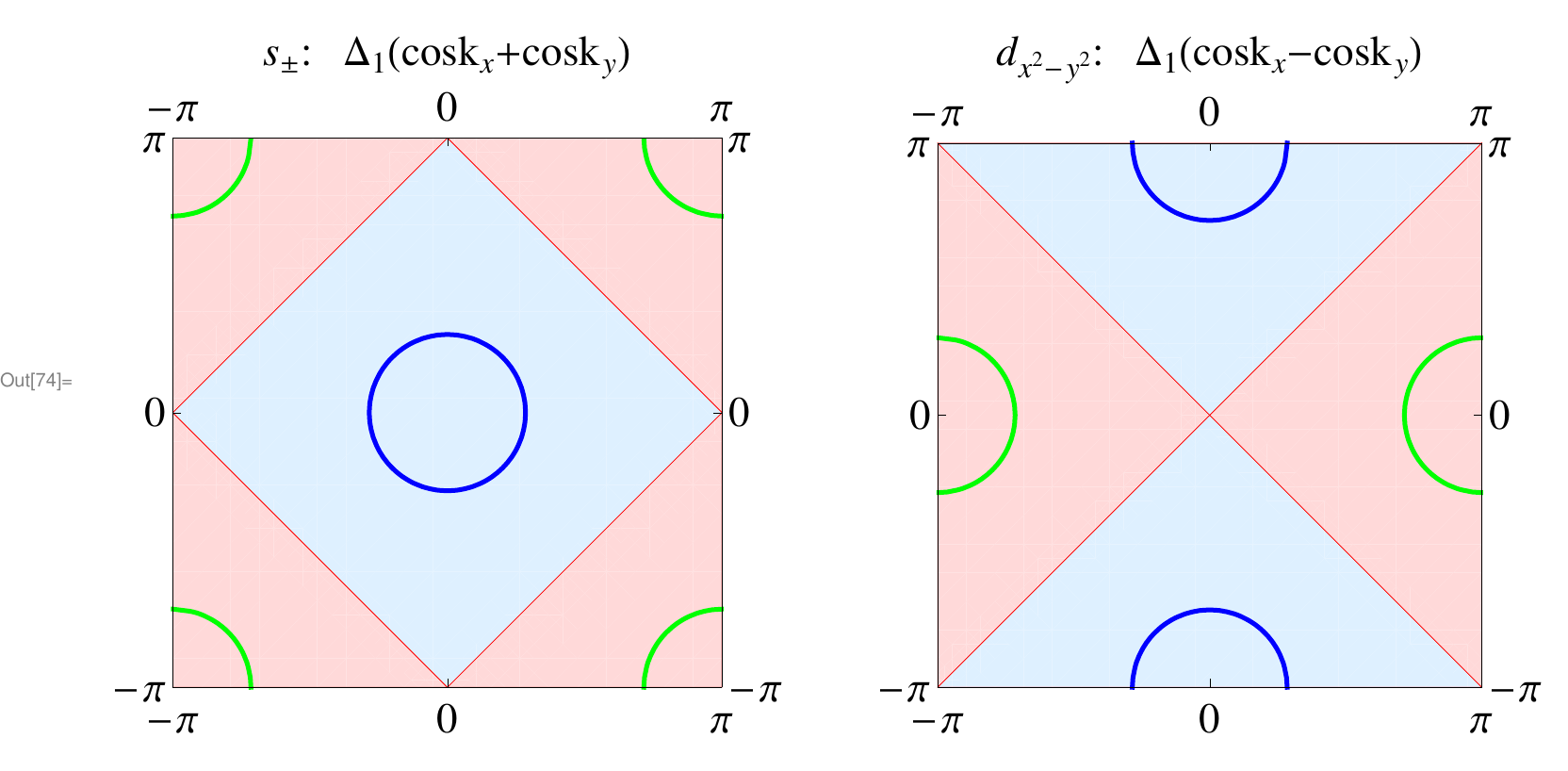}}}
\caption{\label{fig:wti} {In each panel, the blue and green circles denote two Fermi surfaces of the surface state of a weak topological insulator, and the red line separates two regions in which $\Delta_{\bm k}$ has opposite signs.}}
\end{figure}

A nodeless $s_{\pm}$ or $d_{x^2-y^2}$ wave superconductor allows the pair potentials on Fermi surfaces
centered at $(0,0)$ and $(\pi,\pi)$ or at $(0,\pi)$ and $(\pi,0)$ to have opposite signs.
Therefore, in principle it is possible to engineer two dimensional time reversal invariant topological superconductivity in a proximity device that hybridize a weak topological insulator with two surface states and a nodeless $s_{\pm}$ or $d_{x^2-y^2}$ wave superconductor. This two possibilities are sketched in Fig.~\ref{fig:wti}.

\subsection{II. Possibility of using a $s$ wave superconductor}

To elaborate this point, we use Eq.~(4) with $c=0$ and $s$ replaced by $s_z$. $s_z=\pm1$ denote
the top/bottom surface state (or layer) in the case of strong topological insulators [ref.~35] or bilayer Rashba films [ref.~24].
$m s_x\tau_z$ can be understood as a small trivial gap at $k=0$ due to the tunneling between the (top-or-bottom) orbital-pseudospins.
In the weak tunneling limit, the proximity induced pairing potential has the form of
\begin{eqnarray}
T\tau_x=[(t^2_T+t^2_B)/2 +(t^2_T-t^2_B)/2\,s_z+t_T t_B\,s_x]\,\tau_x\,,
\end{eqnarray}
where $t_T$ ($t_B$) is the tunneling amplitude between the superconductor and the top (bottom) orbital.
We have assumed $t_T$ and $t_B$ are real without loss of generality, and will use $T_{I,z,x}$ to denote the three coefficients quadratic in $t$ in the bracket.
Note that only the $s_z\tau_x$ pairing, breaking inversion symmetry, gives rise to the topological superconducting state in class DIII.
However, such a state will compete with the trivial superconducting state.
Focusing on $\mathcal{H}_{SC}$, the superconducting gap is proportional to $|T_{I}-\sqrt{T_{x}^2+T_{z}^2}|$ which vanishes.
Therefore the hybrid superconductor is nodal. Even if the inter-orbital pairing is not allowed,
{\em i.e.}, $T_{x}=0$, $T_{I}>|T_{z}|$ is generally valid and therefore pins the hybrid superconductor to the trivial state.
When the $s$ wave superconductor is integrated out in the standard second order perturbation theory, the renormalized pairing potential reads
\begin{eqnarray}
\widetilde{\Delta}=\frac{\Gamma_N\Delta}{\sqrt{\Delta^2-\omega^2}}\,\mbox{det}^{-1}(1+\frac{\Gamma_N}{\sqrt{\Delta^2-\omega^2}})
\end{eqnarray}
with $\Gamma_N=\pi N_0 VT$.
Here $V$ and $N_0$ are the tunneling volume and the density of states of the superconductor in the normal state at the Fermi energy.
Interestingly and obviously, the conclusions are unchanged at the Fermi energy in this more rigorous approach.
Similar analysis and conclusions can be generalized to the case with both orbital and spin flavors, in which the spin-triplet pairing $\sigma_z s_y \tau_x$ leads to a topological state.